\documentclass{article}
\usepackage{spconf,amsmath,graphicx,amsthm,amssymb,url}
\usepackage{cite} 
\usepackage{multirow}
\usepackage{algorithm}
\usepackage{algpseudocode}
\usepackage{colortbl}  
\usepackage{tabularx}
\usepackage[table,xcdraw]{xcolor}
\usepackage{booktabs}
\usepackage{subcaption}
\usepackage[utf8]{inputenc}
\usepackage[normalem]{ulem}
\usepackage{hyperref}
\usepackage{setspace}

\title{Noise-Aware Speech Separation with Contrastive Learning}
\name{Zizheng Zhang$^1$, Chen Chen$^2$, Hsin-Hung Chen$^3$, Xiang Liu$^1$, Yuchen Hu$^2$, Eng Siong Chng$^2$\thanks{Repository: \href{https://github.com/TzuchengChang/NASS}{https://github.com/TzuchengChang/NASS}}}
\address{
$^1$School of Software and Microelectronics, Peking University, China\\
$^2$School of Computer Science and Engineering, Nanyang Technological University, Singapore\\
$^3$School of Electrical and Computer Engineering, Georgia Institute of Technology, USA\\
$^1$zzz128@stu.pku.edu.cn
}
\begin{document}
%\ninept
%
\maketitle
\begin{abstract}
Recently, speech separation (SS) task has achieved remarkable progress driven by deep learning technique. However, it is still challenging to separate target speech from noisy mixture, as the neural model is vulnerable to assign background noise to each speaker. In this paper, we propose a noise-aware SS (NASS) method, which aims to improve the speech quality for separated signals under noisy conditions. Specifically, NASS views background noise as an additional output and predicts it along with other speakers in a mask-based manner. To effectively denoise, we introduce patch-wise contrastive learning (PCL) between noise and speaker representations from the decoder input and encoder output. PCL loss aims to minimize the mutual information between predicted noise and other speakers at multiple-patch level to suppress the noise information in separated signals. Experimental results show that NASS achieves 1 to 2dB SI-SNRi or SDRi over DPRNN and Sepformer on WHAM! and LibriMix noisy datasets, with less than 0.1M parameter increase.
\end{abstract}
\section{Introduction}
\label{sec:intro}
Speech separation (SS) aims to separate speech signals from overlapping speech mixture\cite{bronkhorst2015cocktail}, which can serve as a pre-processor for downstream speech applications. Recently, deep learning methods have developed various neural networks for SS\cite{luo2019conv, luo2020dual, subakan2021attention, linhui2023monaural, sun2022monaural}, and achieve remarkable performances on clean datasets\cite{hershey2016deep, cosentino2020librimix}. However, it is still challenging to separate target signals from noisy mixture, e.g., \emph{noisy speech separation}, since noise signal usually has a wide distribution on frequency domain to interfere with the human voice. 

For noisy SS, mainstream mask-based methods are vulnerable to assign background noise to target speakers. One intuitive solution is to utilize speech enhancement (SE)\cite{chen2023metric} as a pre-processor to denoise from the mixture in a multi-task manner\cite{ma2021multitask}. Despite the slight improvement, this method may lead to an over-suppression problem\cite{hu2022interactive}, since SE module would inevitably remove some helpful information when denoising, thus resulting in a sub-optimal performance. 

To alleviate the influence of noise, our basic idea is to view background noise as an independent output, which can be predicted along with other speakers simultaneously. In addition to avoiding over-suppression problem, the estimated noise can benefit separated speech from a mutual information\cite{belghazi2018mutual} perspective: We aim to minimize the mutual information between predicted noise and separated speech, thus preventing noise from existing in separated speech. 

In this paper, we propose a noise-aware SS (NASS) method, which follows a typical encoder-separator-decoder pipeline\cite{luo2019conv}. Unlike previous works, NASS learns to predict the noise and leverage it to improve the speech quality of each speaker. Specifically, we conduct patch-wise contrastive learning\cite{park2020contrastive, chen2022noise} between noise and speaker representations from the decoder input and encoder output. For each utterance training, we perform the following loss evaluation: 1) We sample hundreds of times from each representation. 2) For each time, one patch from speech ground-truth are viewed as positive example to predicted speech patch in the same position, while 3) other patches from noise are all viewed as negative examples. Positive and negative examples are calculated with cosine similarity. By solving the classification of positive and negatives, we thus minimize the mutual information between each speaker and noise to significantly suppress the noise from separated speech signals. 

To evaluate the effect of NASS, we conduct intensive experiments on WHAM!\cite{wichern2019wham} and LibriMix\cite{cosentino2020librimix} noisy datasets. We select two milestone works, DPRNN\cite{luo2020dual} and Sepformer\cite{subakan2021attention}, as the separator's backbone. Experimental results show that NASS achieves 1 to 2dB SI-SNRi\cite{le2019sdr} or SDRi\cite{vincent2006performance} over DPRNN and Sepformer on WHAM! and LibriMix noisy datasets, with less than 0.1M parameter increase.

\begin{figure*}[t!]
\centering
\includegraphics[scale=0.8]{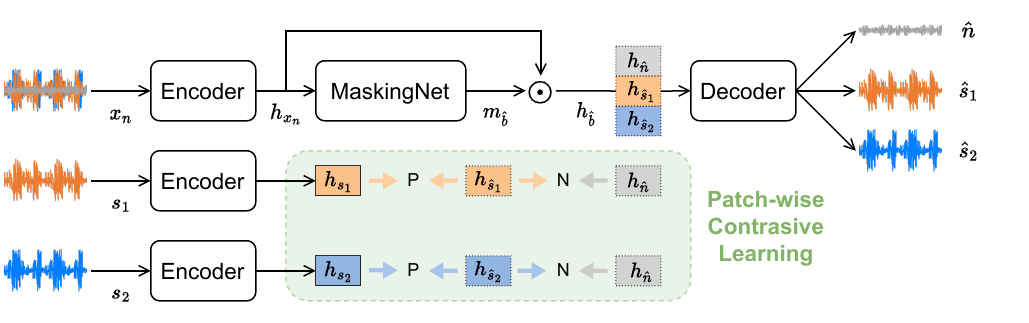}
\vspace{-0.1in}
\caption{The overall pipeline of NASS. $x_n$ and $\hat n$ denote the noisy input and predicted noise. $\hat{s}_1$ and $\hat{s}_2$ are separated speech while $s_1$ and $s_2$ are the ground-truth. $h_{\hat s_1}$, $h_{\hat s_2}$ and $h_{\hat n}$ in dashed box are predicted representations, while $h_{s_1}$ and $h_{s_2}$ in solid box are the ground-truth. ``P" denotes the mutual information between separated and ground-truth speech is maximized while ``N" denotes the mutual information between separated speech and noise is minimized. }
\vspace{-0.1in}
\label{NASS}
\end{figure*}

\section{NASS Method}
We now introduce our proposed NASS method, which consists of mask-based architecture and patch-wise contrastive learning strategy. 

\subsection{Mask-based Architecture}
We follow the encoder-separator-decoder pipeline, as shown in Fig.\ref{NASS}, where NASS is theoretically applicable to any separator or masking net. 

\subsubsection{Encoder and Decoder}
Encoder takes the noisy input ${x_n} \in {\mathbb{R}^{1 \times T}}$ from time domain and generates a STFT-like\cite{wang2018supervised} representation ${h_{x_n}} \in {\mathbb{R}^{F \times L}}$, where $T$ is the signal length, $F$ is the number of filters and $L$ is the number of vectors: 

\vspace{-0.1in}
\begin{equation}
{h_{x_n}}{\rm{ = }}{\mathop{\rm ReLU}\nolimits} ({\mathop{\rm Conv1d}\nolimits} ({x_n}))
\label{eq1}
\end{equation}

Decoder acts as an inverse operation of encoder, which takes the predicted representation ${h_{\hat b}} \in {\mathbb{R}^{F \times L}}$ and reconstructs the separated signal ${y_{\hat b}} \in {\mathbb{R}^{1 \times T}}$ in time domain, where ${b} \in \left\{ {s_1},{s_2}, \ldots ,{s_C}, {n} \right\}$ and $n$ denotes the noise: 

\vspace{-0.1in}
\begin{equation}
{y_{\hat b}} = {\mathop{\rm Conv1dTranspose}\nolimits} ({h_{\hat b}})
\label{eq2}
\end{equation}

In our work, additionally, the ground-truth speech signal ${s_a} \in {\mathbb{R}^{1 \times T}}$ is encoded as ${h_{s_a}} \in {\mathbb{R}^{F \times L}}$, which is utilized along with ${h_{\hat b}}$ in subsequent contrastive learning, where ${a} \in \left\{ {{1},{2}, \ldots ,{C}} \right\}$ and $C$ is the number of speakers. 

\subsubsection{Masking Net and Additional Noise Output}
Though masking net varies from different separators, it takes $h_{x_n}$ and learns a mask ${m_{\hat b}} \in {\mathbb{R}^{F \times L}}$, then yielding $h_{\hat b}$: 

\vspace{-0.1in}
\begin{equation}
{m_{\hat b}} = {\mathop{\rm MaskingNet}\nolimits} ({h_{{x_n}}})
\label{eq3}
\end{equation}

\vspace{-0.1in}
\begin{equation}
{h_{\hat b}} = {m_{\hat b}} \odot {h_{{x_n}}}
\label{eq4}
\end{equation}

In previous work, noise is typically removed directly without being preserved. To make use of the noise information within existing framework. We count noise as an additional output and it works well, which has its own supervision and prediction like speakers. From Eq.\ref{eq2}, \ref{eq3} and \ref{eq4}, we have noise mask ${m_{\hat{n}}} \in {\mathbb{R}^{F \times L}}$, predicted noise representation $h_{\hat n} \in {\mathbb{R}^{F \times L}}$ and predicted noise ${\hat n} \in {\mathbb{R}^{1 \times T}}$. Thus far we have a total of $C + 1$ sources. 

\begin{figure}[t!]
\centering
\includegraphics[width=\linewidth]{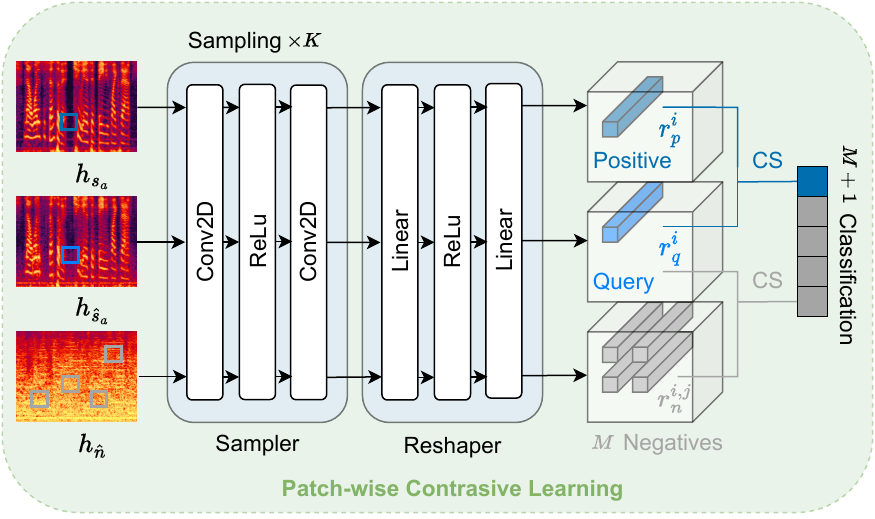}
\vspace{-0.2in}
\caption{The illustration of patch-wise contrastive learning. For the $i$-th sampling of $K$ times, one query example $r^i_q$, positive example $r^i_p$ and $M$ negative examples ${r_n^{i,j}}$ ($j \in [1,M]$) are sampled from predicted speech representation $h_{\hat s_a}$, ground-truth speech representation $h_{s_a}$ and predicted noise representation $h_{\hat n}$, respectively, "CS" denotes cosine similarity. }
\vspace{-0.1in}
\label{Patch}
\end{figure}

\subsection{Patch-wise Contrastive Learning}
Based on additional noise output (ANO), we can further denoise the separated speech by performing patch-wise contrastive learning (PCL) between $h_{\hat s_a}$, $h_{s_a}$ and $h_{\hat n}$, as shown in Fig.\ref{Patch}. We now describe the contrastive learning process for each utterance training.

\subsubsection{Sampler and Reshaper}
First, sampler would sample $K$ times on $h_{\hat s_a}$, $h_{s_a}$ and $h_{\hat n}$. For each time, $1$, $1$, $M$ patches are randomly sampled from $h_{\hat s_a}$, $h_{s_a}$, $h_{\hat n}$ and viewed as query, positive, negatives examples, respectively, where positive, query and one of the negative examples are in the same position while $M - 1$ negatives are different. Since sampler consists of two convolutional layer with a kernel size of $H$, the patch-size is $H^2$:

\vspace{-0.1in}
\begin{equation}
{r_p^i,r_q^i,r_n^{i,j}}{\rm{ = }}{\mathop{\rm Conv2d}\nolimits} ({\mathop{\rm ReLU}\nolimits} ({\mathop{\rm Conv2d}\nolimits} ({h_{{s_a}}},{h_{{{\hat s}_a}}},{h_{\hat n}})))
\label{equ5}
\end{equation}

where ${r_p^i,r_q^i,r_n^{i,j}} \in {\mathbb{R}^{H^2 \times H^2}}$ ($i \in [1,K]$, $j \in [1,M]$) are the positive, query and negative examples in the $i$-th sampling from ${h_{{s_a}}}$, ${h_{{{\hat s}_a}}}$, ${h_{\hat n}}$, respectively.

Then reshaper learns to project all query, positive and negative examples into a 3-D embedding space for contrastive learning, which consists of a two-layer MLP: 

\vspace{-0.1in}
\begin{equation}
{r_p,r_q,r_n^j}{\rm{ = }}{\mathop{\rm Linear}\nolimits} ({\mathop{\rm ReLU}\nolimits} ({\mathop{\rm Linear}\nolimits} ({r_p^i,r_q^i,r_n^{i,j}})))
\label{equ6}
\end{equation}

where ${r_p,r_q,r_n^j} \in {\mathbb{R}^{H^2 \times H^2 \times K}}$. In this work, $H$, $K$, $M$ is set to $3$, $256$, $256$, respectively.

\subsubsection{Contrastive Loss}
From a perspective of mutual information\cite{belghazi2018mutual}, query example should be similar to corresponding positive example while different to all negative examples, which provides an $M + 1$ classification. Through gradient passing, separator will learn to minimize the mutual information between query and negative examples, thus suppressing the noise from separated speech. Contrastive loss is conducted from each of $C$ speakers by calculating the cosine similarity and optimizing the cross-entropy loss, which can be formulated as: 

\vspace{-0.1in}
\begin{equation}
\mathcal{L}_{pcl} = \frac{{\rm{1}}}{C}\sum\limits_{t = 1}^C {\sum\limits_{i = 1}^M { - \ln } } \left[ {\frac{{{e^{r_q^i \cdot r_p^i/\tau }}}}{{{e^{r_q^i \cdot r_p^i/\tau }} + \sum\nolimits_{j = 1}^M {{e^{r_q^i \cdot r_n^{i,j}/\tau }}} }}} \right]
\label{equ7}
\end{equation}

where  ``$\cdot$" denotes the computation of cosine similarity, $\tau$ denotes the temperature parameter, which is set to 0.07.

\subsection{Training Objective}
Main loss $\mathcal{L}_{si-snr}$ is to maximize SI-SNR\cite{le2019sdr} between separated signal ${y_{\hat b}}$ and ground-truth signal ${y_b}$ for $C + 1$ sources: 

\vspace{-0.1in}
\begin{equation}
\mathcal{L}_{si-snr} = \frac{1}{C+1}\sum\limits_{b = 1}^{C+1} { - 10{{\log }_{10}}\left( {\frac{{{{\Vert {\frac{{{{y_{\hat b}}^T}{y_b}}}{{{{\Vert {{y_b}} \Vert}^2}}}{y_b}} \Vert}^2}}}{{{{\Vert {\frac{{{{y_{\hat b}}^T}{y_b}}}{{{{\Vert {{y_b}} \Vert}^2}}}{y_b} - {{y_{\hat b}}}} \Vert}^2}}}} \right)}
\label{equ8}
\end{equation}

Thus total loss of proposed NASS is formulated as: 

\vspace{-0.1in}
\begin{equation}
\mathcal{L}_{total} = \mathcal{L}_{si-snr} + \lambda \mathcal{L}_{pcl}
\label{equ9}
\end{equation}

where $\lambda$ is the parameter to balance $\mathcal{L}_{si-snr}$ and $\mathcal{L}_{pcl}$, which is set to 2 in this work.

NASS is trained with uPIT\cite{kolbaek2017multitalker}. It is worth noting that $\mathcal{L}_{pcl}$ depends on the permutation result of uPIT, which reduces double-counting and determines the order of output for corresponding comparisons.

\begin{table}[t]
\caption{Ablation study of additional noise output (ANO) on noisy LibriMix. Baseline results with ``$^*$" are self-reproduced.}
\centering
\resizebox{\linewidth}{!}{
\begin{tabular}{c|c|c|c|c|c}
\toprule[1.5pt]
Method & Spk & ANO & SI-SNRi (dB) & SDRi (dB) & Params (M)\\
\midrule
\multirow{4}{*}{DPRNN$^*$} & \multirow{2}{*}{2} & $\times$ & 13.1 & 13.7 & 14.8\\
\multirow{4}{*}{} & \multirow{2}{*}{} & \checkmark & \textbf{13.2} & \textbf{13.8} & 14.8\\
\cmidrule{2-6}
\multirow{4}{*}{} & \multirow{2}{*}{3} & $\times$ & 11.3& 11.9 & 14.9\\
\multirow{4}{*}{} & \multirow{2}{*}{} & \checkmark & \textbf{11.9} & \textbf{12.4} & 14.9\\
\midrule
\multirow{4}{*}{Sepformer$^*$} & \multirow{2}{*}{2} & $\times$ & 13.2 & 13.8 & 25.8\\
\multirow{4}{*}{} & \multirow{2}{*}{} & \checkmark & \textbf{13.5} & \textbf{14.1} & 25.8\\
\cmidrule{2-6}
\multirow{4}{*}{} & \multirow{2}{*}{3} & $\times$ & 10.0 & 10.5 & 25.9\\
\multirow{4}{*}{} & \multirow{2}{*}{} & \checkmark & \textbf{11.0} & \textbf{11.6} & 25.9\\
\bottomrule[1.5pt]
\end{tabular}}
\label{ano}
\end{table}
\vspace{-0.1in}

\begin{table}[t]
\caption{Ablation study of patch-wise contrastive learning (PCL) on noisy Libri2Mix with Sepformer. ``$s\xrightarrow{}n$" denotes $\mathcal{L}^{s\xrightarrow{}n}_{pcl}$ while ``$n\xrightarrow{}s$" denotes $\mathcal{L}^{n\xrightarrow{}s}_{pcl}$. ``$\lambda$" is the balancing parameter and ``$M$" is the number of negative patches.}
\centering
\resizebox{\linewidth}{!}{
\begin{tabular}{c|c|c|c|c|c|c}
\toprule[1.5pt]
$s\xrightarrow{}n$ & $n\xrightarrow{}s$ & $\lambda$ & $M$ &SI-SNRi (dB) & SDRi (dB) & Params (M)\\
\midrule
$\times$ & $\times$ & - & - & 13.5 & 14.1 & 25.8\\
$\times$ & \checkmark & 1 & 256 & 13.9 & 14.4 & 25.8\\
\checkmark & $\times$ & 1 & 256 & 14.0 & 14.6 & 25.8\\
\checkmark & $\times$ & 1 & 1 & 13.6 & 14.2 & 25.8\\
\checkmark & $\times$ & 2 & 256 & \textbf{14.4} & \textbf{15.0} & 25.8\\
\bottomrule[1.5pt]
\end{tabular}}
\vspace{-0.1in}
\label{pcl}
\end{table}

\section{Experiments}
\subsection{Datasets}
We evaluate NASS and existing methods on two noisy datasets: WHAM!\cite{wichern2019wham} and LibriMix\cite{cosentino2020librimix}.
WHAM! is a noisy version of WSJ0-2Mix, which is added noise samples recorded in coffee shops, restaurants and bars. SNR between the loudest speaker and noise varies from -6 to +3 dB. WHAM! follows the same structure as WSJ0-2Mix, which has 119 speakers and 43 hours of speech. LibriMix is for multi-speaker noisy SS tasks. In our chosen of LibriMix, clean mixture is selected from LibriSpeech train-100 and mixed between -25 and -33 dB of LUFS. Noise samples from WHAM! are added to the mixture between -38 and -30 dB of LUFS. With 331 speakers, Libri2Mix has 80 hours of speech while Libri3Mix has 62 hours. 

\subsection{Experimental Setup}
To ensure reproducibility, we conduct experiments on SpeechBrain\cite{ravanelli2021speechbrain} and provide a \href{https://github.com/TzuchengChang/NASS}{github repository}, where configurations and more details can be found. 

We train 200 epochs for all experiments on NVIDIA V100 GPUs, using Adam optimizer with initial learning rate of $1.5\times10^{-4}$ and automatic mixed precision. After training 50 epochs, learning rate would be halved if with no improvement of validation for 3 epochs. Speed perturbation is applied for data augmentation. There's no dynamic mixing in our experiments. Batch size and number of workers are set to 1 and 4. Training signal length is 4 seconds long and loss threshold is set to -30. Gradient clipping is applied to limit the $L_2$ norm of gradients to 5.

\subsection{Metrics and Baselines}
We use SI-SNRi\cite{le2019sdr} and SDRi\cite{vincent2006performance} from test-set to evaluate all experiments, which measure the degree of improvement in clarity and fidelity of separated signals. For all metrics, higher score indicates better performance.

To assess the effectiveness of NASS, we select DPRNN\cite{luo2020dual} and Sepformer\cite{subakan2021attention} as baselines for comparisons.

\begin{figure}[t!]
\centering
\includegraphics[width=\linewidth]{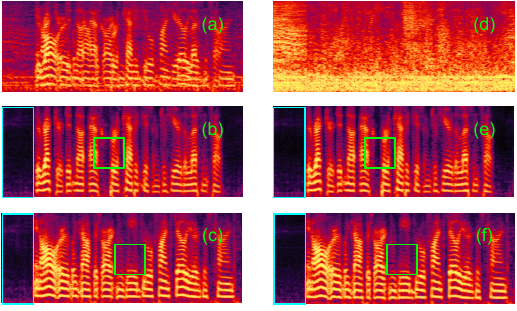}
\vspace{-0.2in}
\caption{Spectrum results on Libri2mix with Sepformer. Subplot (a) denotes the mixture; (b), (c) are baseline results; (d), (e), (f) are NASS results. Note that (d) is the noise output.}
\vspace{-0.1in}
\label{Feature}
\end{figure}

\section{Results}
\subsection{Effect of Additional Noise Output}
As shown in Table\ref{ano}, compared to baseline results, benefit from noise supervision, methods with additional noise output (ANO) perform better, especially in the case of 3 speakers. Note that noise loss is counted during training while test results from SI-SNRi and SDRi include speakers only.

\subsection{Effect of Patch-wise Contrastive Learning}
Based on ANO, we study the effect of patch-wise contrastive learning (PCL) in Table\ref{pcl}. As mentioned in Eq.\ref{equ7}, query example $r_q^i$ and negative examples $r_n^{i,j}$ are from speech and noise, respectively. In this case, loss is denoted as $\mathcal{L}^{s\xrightarrow{}n}_{pcl}$. However, query example can be from noise and vice versa, denoting the loss as $\mathcal{L}^{n\xrightarrow{}s}_{pcl}$. Results show that the best choice is $\mathcal{L}^{s\xrightarrow{}n}_{pcl}$, and the best balancing parameter $\lambda$ is $2$ instead of $1$, indicating that PCL works on ANO, and the number of negative patches $M$ are best set to $256$ instead of $1$, also indicating that why we choose more patches on noise representation, since if $M$ is $1$, it will regrades to a binary-classification, which not as much informative.

We don't perform PCL between speakers, since mutual information (MI) between them is indeterminate, e.g., \emph{they may have very similar voices}, while MI between speakers and noise is definitely in need of reduction.

\subsection{Competitive Results of NASS}
Table\ref{librimix} and Table\ref{wham} show NASS results on noisy LibriMix and WHAM!. They indicate that NASS could make previous models like DPRNN and Sepformer surpass themselves under noisy condition, with less than 0.1M additional parameters. In addition, we find NASS also works to ConvTasnet\cite{luo2019conv}, which is not shown for space constraints, and it may apply to other non-mask-based models as well.

To further illustrate the effect of NASS, Fig.\ref{Feature} visualizes the results of spectrum. We can see some noise spectrum in (b), (c) from Sepformer baseline fades away in a degree to (e), (f) from NASS Sepformer, thus indicating better separation. Besides, we can see NASS yields the prediction of noise in (d), which can be used in other speech tasks.v

\begin{table}[t!]
\caption{Competitive results of proposed NASS on noisy LibriMix. Baseline results with ``$^*$" are self-reproduced.}
\centering
\resizebox{\linewidth}{!}{
\begin{tabular}{c|c|c|c|c}
\toprule[1.5pt]
Method & Spk & SI-SNRi (dB) & SDRi (dB) & Params (M)\\
\midrule
\multirow{2}{*}{DPRNN$^*$} & 2 & 13.1 & 13.7 & 14.8\\
\multirow{2}{*}{} & 3 & 11.3 & 11.9 & 14.9\\
\cmidrule{2-5}
\multirow{2}{*}{Sepformer$^*$} & 2 & 13.2 & 13.8 & 25.8\\
\multirow{2}{*}{} & 3 & 10.0 & 10.5 & 25.9\\
\midrule
\multirow{2}{*}{DPRNN (NASS)} & 2 & \textbf{13.5} & \textbf{14.1} & 14.8\\
\multirow{2}{*}{} & 3 & \textbf{12.4} & \textbf{12.9} & 14.9\\
\cmidrule{2-5}
\multirow{2}{*}{Sepformer (NASS)} & 2 & \textbf{14.4} & \textbf{15.0} & 25.8\\
\multirow{2}{*}{} & 3 & \textbf{12.1} & \textbf{12.7} & 25.9\\
\bottomrule[1.5pt]
\end{tabular}}
\label{librimix}
\end{table}
\vspace{-0.1in}

\begin{table}[t!]
\caption{Competitive results of proposed NASS on WHAM!. Note that WHAM! has only 2-spk version. Baseline results are reported from the original paper.}
\centering
\resizebox{\linewidth}{!}{
\begin{tabular}{c|c|c|c}
\toprule[1.5pt]
Method & SI-SNRi (dB) & SDRi (dB) & Params (M)\\
\midrule
DPRNN\cite{luo2020dual} & 13.7 & 14.1 & 2.7\\
Sepformer\cite{subakan2021attention} & 14.4 & 15.0 & 26.0\\
\midrule
DPRNN (NASS) & \textbf{15.8} & \textbf{16.1} & 14.9\\
Sepformer (NASS) & \textbf{16.5} & \textbf{16.8} & 25.8\\
\bottomrule[1.5pt]
\end{tabular}}
\vspace{-0.1in}
\label{wham}
\end{table}

\section{Conclusions}
In this work, we propose a noise-aware speech separation (NASS) method. Specifically, we set up an additional noise output (ANO) to make use of the noise information. Based on ANO, patch-wise contrastive learning (PCL) is conducted to further help separate sources in detail. Experimental results show that NASS achieves 1 to 2dB SI-SNRi or SDRi over DPRNN and Sepformer on WHAM! and LibriMix noisy datasets, with less than 0.1M parameter increase.

\clearpage

% References should be produced using the bibtex program from suitable
% BiBTeX files (here: strings, refs, manuals). The IEEEbib.bst bibliography
% style file from IEEE produces unsorted bibliography list.
% -------------------------------------------------------------------------
\begin{spacing}{0.9}
\bibliographystyle{IEEEbib}
\bibliography{refs}
\end{spacing}

\end{document}